\documentstyle[12pt,psfig]{article}
\begin{document}

\hfill PUPT-1786

\hfill hep-th/9804075

\vspace{1.5in}

\begin{center}

{\large\bf Equivariant Sheaves}\footnote{Invited paper
to appear in the special issue of the journal {\it Chaos, Solitons,
and Fractals \/} on ``Superstrings, M, F, S, ... Theory'' (M.S. El
Naschie and C. Castro, editors).}

\vspace{0.5in}

Allen Knutson \\
Mathematics Department \\
Brandeis University \\
Waltham, MA  02254 \\
{\tt allenk@alumni.caltech.edu} \\

\hspace{0.5in}  \\

Eric Sharpe \\
Physics Department \\
Princeton University \\
Princeton, NJ  08544 \\
{\tt ersharpe@puhep1.princeton.edu}

\hspace{0.5in} \\

\end{center}

In this article we review some recent developments in
heterotic compactifications.   In particular we review
an ``inherently toric'' description of certain sheaves,
called equivariant sheaves, that has recently been discussed
in the physics literature.  We outline calculations that can
be performed with these objects, and also outline more general
phenomena in moduli spaces of sheaves.

\begin{flushleft}
April 1998
\end{flushleft}

\newpage

\section{Introduction}

Twelve years ago, during the first superstring revolution,
there was virtually no technology at all to describe compactifications.
Since then, matters have improved greatly.

For some string theories, namely the type II theories and M and
F theory, a compactification requires essentially\footnote{
We are ignoring, for example, flat M theory 3-form potentials
which can be turned on.}
specifying
only a Calabi-Yau.  These compactifications are now reasonably
well understood.  Not only do we have technology for analyzing
many Calabi-Yaus, but we also have a basic understanding of
quantum effects, both in $\alpha'$ (such as mirror
symmetry) and in the string coupling
constant (such as enhanced gauge symmetries arising from
singular Calabi-Yaus).

There are additional string theories (the heterotic and
type I theories) whose compactifications are unfortunately
understood much more poorly.  The complication in these
cases is that to compactify one must specify not only
a Calabi-Yau, but also at least one bundle (or, more
generally, a torsion-free sheaf) over the Calabi-Yau.
Although physicists now have a lot of technology to analyze
Calabi-Yaus, there are relatively few ways to get any handle
on bundles on Calabi-Yaus.

Until recently, there were only two known ways to describe\footnote{
In a manner that gives one control over moduli.  Toroidal orbifolds
and WZW models, for example, can also be used to construct heterotic
compactifications, but with no handle on continuous moduli.
Studying an entire moduli space by studying a few points is much
like studying an ocean by studying a few water molecules -- one
will learn much about water, but nothing about waves, fish, or
most of the other things that make an ocean interesting.}
sheaves on Calabi-Yaus.  First, about a year ago a description
of bundles on elliptic Calabi-Yaus was published \cite{fmw}.
Their description is quite beautiful, but unfortunately
only describes bundles, not more general sheaves, and only
on elliptic Calabi-Yaus, not more general Calabi-Yaus.
The other description of sheaves on Calabi-Yaus was first
published several years prior \cite{02}.  The Distler-Kachru
models described therein give one excellent control over the
physics of heterotic compactifications, but are extremely
cumbersome to work with mathematically.  

In this article
we shall describe a third approach to the problem of
describing sheaves on Calabi-Yaus.
Specifically, we shall review a very convenient set of sheaves
on toric varieties -- ``equivariant'' sheaves -- which are
mathematically quite easy to work with.  Most of the Calabi-Yaus
studied by physicists are realized as hypersurfaces
(or complete intersections) in toric varieties, so sheaves on Calabi-Yaus
can be constructed by restriction of a sheaf on a toric variety
to the Calabi-Yau
hypersurface (or complete intersection).  (Not all sheaves on Calabi-Yaus
can be constructed this way, however we can obtain a large subfamily.)
Unfortunately we will
have nothing to say about worldsheet instanton corrections
 -- our discussion will be purely classical in nature.
Equivariant sheaves were recently discussed by the authors
in \cite{meallen}, in work which built upon the prior
work largely by A. A. Klyachko 
\cite{klyachko1,klyachko2,klyachko3,klyachko4,kaneyama1,kaneyama2}.

Why is it important to understand heterotic\footnote{Or, for
that matter, type I compactifications.  Type I compactifications
have additional technical complications beyond those of
heterotic compactifications, so for this article we shall
only be concerned with heterotic compactifications.}
compactifications?  For physicists, there are several good
reasons.  First, even after the advent of string duality,
heterotic compactifications remain the most technically efficient
ways to get phenomenologically viable results.
Secondly, via string duality a good understanding of quantum
effects in heterotic theories would surely yield insight
into compactifications of other string theories.

For mathematicians, heterotic compactifications are interesting
because of the potential existence of a generalization of mirror
symmetry known as (0,2) mirror symmetry, which we shall
discuss at greater length in section~\ref{02mirrsec}.

In section~\ref{tv} we shall begin with a brief overview
of some relevant characteristics of toric varieties.
In section~\ref{eb} we shall describe equivariant bundles
on toric varieties.  In section~\ref{es} we shall describe
more general equivariant torsion-free sheaves on toric varieties,
and outline the origins of the description presented herein.
In section~\ref{modspace} we shall discuss the construction
of moduli spaces of equivariant sheaves, and some of their
prominent characteristics.  In section~\ref{genmodspace} we shall
comment on more general moduli spaces of sheaves.  Finally
in section~\ref{02mirrsec} we shall comment on (0,2) mirror
symmetry.

\section{Toric varieties}  \label{tv}

Why are physicists interested in toric varieties?
Essentially because most of the Calabi-Yaus presently
studied are realized as hypersurfaces (or complete intersections)
in toric varieties.  Toric varieties are reasonably
well-understood, in the sense that most computations one
would like to perform are relatively straightforward.

What is a ``toric variety''?  A toric variety\footnote{
For more information on toric varieties see \cite{fulton,oda,kkmsd,danilov}.  
Not all
compactifications of algebraic tori are toric varieties -- toric
varieties have additional nice properties -- but the distinctions
will not be relevant for the purposes of this article.} is, for the
purposes of this paper, a variety which is an at least partial 
compactification
of an ``algebraic torus'' -- a product of ${\bf C}^{\times} = 
{\bf C} - \{ 0 \}$ 's -- such that the algebraic torus action
extends continuously over the entire variety.  
(Each ${\bf C}^{\times}$ contains
an $S^1$, so an algebraic torus can be thought of as a sort
of complexification of an ordinary torus, thus the name.)  For example, 
\begin{displaymath}
{\bf P}^1 \: = \: {\bf C}^{\times}
\cup \{ 0 \} \cup \{ \infty \}
\end{displaymath}
Another example is
\begin{displaymath}
{\bf P}^2 \: = \: ( {\bf C}^{\times} )^2 \cup \{ x = 0 \}
\cup \{ y = 0 \} \cup \{ z = 0 \}
\end{displaymath}
where $x$, $y$, and $z$ are homogeneous coordinates defining the
toric variety. 

Toric varieties often have a description in terms of
homogeneous coordinates \cite{cox}.
How can the algebraic torus be seen in such a description?
The algebraic torus is simply all possible ${\bf C}^{\times}$
rescalings of the individual homogeneous coordinates, modulo
the ${\bf C}^{\times}$'s one mods out to form the toric variety.

The codimension one subvarieties that compactify
the algebraic torus are called ``toric divisors.''
In the ${\bf P}^2$ example, the sets $\{ x= 0 \}$, $\{ y = 0 \}$,
and $\{ z = 0 \}$ are the toric divisors.

In general, if we know everything about how the toric divisors
are attached,
then we know almost everything about the toric variety.
Loosely speaking, given knowledge of the toric divisors we
can use the underlying algebraic torus $( {\bf C}^{\times} )^n$
to sweep out the rest of the toric variety.

\section{Equivariant bundles}  \label{eb}

Given that all toric varieties are a compactification of an algebraic
torus $( {\bf C}^{\times} )^n$, what can we say about
bundles on toric varieties?

Let $t \in ( {\bf C}^{\times} )^n$, so $t$ has a natural action
on the toric variety -- $t$ simply rotates the underlying algebraic
torus.  (On ${\bf P}^1$, for example, this would correspond
to rotations about an axis plus dilations that leave two poles
fixed.)

Now, given any bundle ${\cal E}$, we can form the bundle
$t^* {\cal E}$ -- we drag ${\cal E}$ back along the action of $t$.
In general, ${\cal E} \not\cong t^* {\cal E}$.

In the special case that ${\cal E} \cong t^* {\cal E}$ for all $t$,
we say that ${\cal E}$ is an equivariant\footnote{In the mathematics
literature, by an equivariant bundle one would typically
mean not only that ${\cal E} \cong t^* {\cal E}$, but one
would have in mind a fixed choice of isomorphisms -- an
``equivariant structure.''  The sheaves we describe in this paper
all are implicitly associated with a specific
choice of equivariant structure, a fact we will return to later.
In this article we shall be somewhat
loose and often ignore the equivariant structure. } 
bundle.
(Equivariant with respect to the underlying algebraic torus.)

It is equivariant bundles (and more generally, equivariant sheaves)
for which there exists a nice description.

What are some examples of equivariant bundles on smooth compact
toric varieties?  First, line bundles are equivariant.
It turns out that all smooth compact toric varieties are simply
connected, so line bundles have no moduli.  Since line bundles
cannot be deformed at all, they certainly cannot be deformed
by the algebraic torus -- thus, line bundles are equivariant.
Similarly, direct sums of line bundles are equivariant.
The tangent and cotangent bundles of a toric variety are
equivariant.  Examples of such bundles are not uncommon,
and typically come in continuous families -- they can certainly
have moduli.

Now, suppose ${\cal E}$ is an equivariant bundle.
It turns out that to reconstruct ${\cal E}$ it suffices to know
its behavior in neighborhoods of the toric divisors.
Given knowledge of ${\cal E}$ near the toric divisors, we can
then (loosely speaking) use the underlying algebraic torus
to rotate that information around and recreate ${\cal E}$ on the
rest of the toric variety.
So, precisely what information must we associate to each toric divisor
to specify an equivariant bundle?  

It turns out that an equivariant bundle ${\cal E}$ can be specified
by associating a ``filtration'' of a vector space to each
toric divisor.  Recall that a filtration of a vector space $E$ is
simply a nested set of vector subspaces of $E$:
\begin{displaymath}
E \supseteq \cdots \supseteq E^{\alpha}(i) \supseteq E^{\alpha}(i+1)
\supseteq E^{\alpha}(i+2) \supseteq \cdots \supseteq 0
\end{displaymath}
The vector space we filter
is precisely the fiber of the vector bundle.

A random set of filtrations does not necessarily define a bundle
 -- they must satisfy a compatibility condition.
On a smooth toric variety this compatibility condition says that 
in any cone of the fan defining the toric variety, 
all the elements of the filtrations
associated to toric divisors in the cone must be coordinate
subspaces of the vector space $E$, with respect to some 
basis of the vector space.  This compatibility 
condition is trivial for two dimensional varieties\footnote{
It turns out that this is the equivariant version of the
statement that on a smooth variety, reflexive sheaves are locally
free up to codimension three.}.  

Two sets of compatible filtrations define the same bundle
precisely when they differ by an automorphism of the
vector space $E$.

Before we can describe some examples, we must first
clear up some loose ends.  The filtration description of
equivariant sheaves given above hinges on a choice of
``equivariant structure'' of the bundle. 
What is an equivariant structure?  We have mentioned that
a bundle ${\cal E}$ is equivariant precisely when it is
isomorphic to $t^* {\cal E}$ for all $t \in ( {\bf C}^{\times} )^n$;
an equivariant structure is simply a precise choice of
isomorphism for each $t$.

The choice of equivariant structure is not unique, and different
choices yield distinct filtrations, but for all that the
choice is relatively harmless -- it adds no continuous
moduli, and is well understood.

Let us consider an example -- line bundles on ${\bf P}^2$.
Let $x$, $y$, and $z$ denote homogeneous coordinates defining the
toric variety, let $D_x$ denote $\{ x = 0 \}$, and so forth.
For readers not acquainted with the notation,
${\cal O}(a)$ denotes a line bundle of $c_1 = a$.

In this context, consider the line bundles
${\cal O}(D_x)$, ${\cal O}(D_y)$, ${\cal O}(3 D_z - 2 D_y)$,
and ${\cal O}(6D_x + 7 D_y - 12 D_z)$.  These line bundles
are all isomorphic as line bundles to ${\cal O}(1)$,
however they all have distinct equivariant structures\footnote{
Note that if we worked with Chern classes in equivariant cohomology
rather than in singular cohomology, we would be able to distinguish
the Chern classes of line bundles with distinct equivariant
structures.}.
More generally, the equivariant structure of a line
bundle is given by a specific choice of torus-invariant
divisor.

How can we describe a line bundle with filtrations?
Consider the example ${\cal O}(n D_x)$ on ${\bf P}^2$.
This line bundle is specified by the filtrations
\begin{eqnarray*}
E^x(i) & = & \left\{ \begin{array}{ll}
                     {\bf C} & i \leq n \\
                     0       & i > n
                     \end{array} \right. \\
E^y(i) \: = \: E^z(i) & = & \left\{ \begin{array}{ll}
                              {\bf C} & i \leq 0 \\
                              0       & i > 0
                              \end{array} \right.
\end{eqnarray*}
Since ${\cal O}(n D_x)$ is a line bundle, its fiber is ${\bf C}$,
so the top vector space in each filtration is ${\bf C}$.
The only complex vector subspace of ${\bf C}$ is $0$,
so each filtration necessarily looks like a string of ${\bf C}$'s
followed by a string of $0$'s.  All information is contained in the
precise value of $i$ at which the filtration changes dimension.
Clearly the filtration description is overkill for line bundles,
but for higher rank bundles it is quite useful.

It turns out that Chern classes and sheaf cohomology groups
of equivariant bundles are quite straightforward to calculate.
We shall not work through the details here, but shall merely
outline the highlights.
For example, if we define
\begin{displaymath}
E^{[\alpha]}(i) \: = \: \frac{ E^{\alpha}(i) }{ E^{\alpha}(i+1) }
\end{displaymath}
then it can be shown that for any bundle ${\cal E}$,
\begin{displaymath}
c_1({\cal E}) \: = \: \sum_{\alpha, i} i \, \mbox{dim }E^{[\alpha]}(i)
\, D_{\alpha}
\end{displaymath}
Sheaf cohomology groups of equivariant bundles have a natural
decomposition, known as an ``isotypic decomposition'',
into subgroups each of which is associated with an element of
the weight lattice of the algebraic torus:
\begin{displaymath}
H^p({\cal E}) \: = \: \bigoplus_{\chi} H^p({\cal E})_{\chi}
\end{displaymath}
Sheaf cohomology of equivariant bundles on smooth
toric varieties can be calculated as 
\u{C}ech cohomology on a natural Leray cover, a straightforward exercise.

\section{Equivariant sheaves}  \label{es}

The rather compact description of equivariant bundles
given above can be generalized to equivariant sheaves.
Before we do so, however, we shall review some basic
definitions.

A locally free sheaf is precisely the sheaf of sections of some
vector bundle.  Each stalk of the sheaf is a freely generated
module, thus the nomenclature.  In this article we shall
fail to distinguish ``bundle'' from ``locally free sheaf.''

A reflexive sheaf is a sheaf ${\cal E}$ such that
${\cal E} \cong {\cal E}^{\vee \vee}$, where ${\cal E}^{\vee}$
is the dual\footnote{For example, the dual of a line bundle
${\cal O}(D)$ is ${\cal O}(-D)$.  In the physics literature
it is traditional to use $*$ rather than $\vee$ to denote
duals; here we follow the notation of algebraic geometers.}
sheaf:  ${\cal E}^{\vee} = {\it Hom}({\cal E}, {\cal O})$.
For example, all bundles are reflexive sheaves.
On a smooth variety, reflexive sheaves are locally free up to 
codimension three.  Also, on a smooth variety all reflexive
rank 1 sheaves are locally free.

A torsion-free sheaf is a sheaf such that each stalk is a 
torsion-free module.  On a smooth variety, torsion-free
sheaves are locally free up to codimension two.
Physicists may think (rather loosely) of torsion-free sheaves
as being bundles with possible small instanton singularities.
In general, reflexive sheaves are special cases of torsion-free sheaves.

Any equivariant torsion-free sheaf looks like a trivial
vector bundle over the open torus orbit.

It turns out that an equivariant reflexive sheaf can be 
specified by associating a filtration\footnote{A filtration
of the fiber of the trivial vector bundle over the 
open torus orbit.} to each toric
divisor.  The difference between an equivariant reflexive
sheaf and an equivariant bundle is that for an equivariant
reflexive sheaf, the filtrations are not required to satisfy
a compatibility condition.

How is this description derived, and how can we describe
more general equivariant torsion-free sheaves?
To explain these matters, we must make a very short digression
into modern algebraic geometry.

Instead of working with topological spaces directly, algebraic
geometers work with rings of functions on spaces.  More precisely,
given any (commutative) ring (with identity), say $A$,
there is a map (Spec) that associates an affine space to $A$:
\begin{displaymath}
\mbox{Spec}: \: \mbox{Rings} \: \rightarrow \: \mbox{Affine spaces}
\end{displaymath}
(One can then build up a compact space by working on coordinate
patches.)  For example,
\begin{eqnarray*}
\mbox{Spec } {\bf C}[x_1, \cdots, x_n] & = & {\bf C}^n \\
\mbox{Spec } {\bf C}[x_1, \cdots, x_n]/(p) & = &
\mbox{the hypersurface } \{ p = 0 \} \subset {\bf C}^n \\
\mbox{Spec } {\bf C} & = & \mbox{a single point}
\end{eqnarray*}

Coherent sheaves over an affine space are described
in terms of modules of sections of the sheaf -- to each
ring $A$, we associate an $A$-module $M$.  Put another way,
we can either speak of pairs (ring $A$, $A$-module $M$)
or of pairs (affine space $U$, sheaf on $U$) = 
(Spec $A$, $\tilde{M}$).  These two descriptions are equivalent!

For example, consider ${\bf C}^2$ and sheaves on ${\bf C}^2$.
The affine space ${\bf C}^2$ is associated with the
polynomial ring ${\bf C}[x,y]$, i.e.,
\begin{displaymath}
\mbox{Spec } {\bf C}[x,y] \: = \: {\bf C}^2
\end{displaymath}
Sheaves on ${\bf C}^2$ are then associated to ${\bf C}[x,y]$-modules.
For example, the trivial rank $r$ veector bundle on ${\bf C}^2$
is associated to the ${\bf C}[x,y]$ module
\begin{displaymath}
\bigoplus_r {\bf C}[x,y]
\end{displaymath}
Note this module is freely generated -- that is why the corresponding
sheaf is locally free.

The description of equivariant torsion-free sheaves can be derived
by thinking along these lines.  First, we shall set up
the sheaf.
To each maximal cone $\sigma$ of the fan defining the toric variety,
associated a ${\bf C}[\sigma^{\vee}]$-module, call it $E^{\sigma}$.
If $\tau$ is a subcone of $\sigma$, then $E^{\tau}$ is defined
to be the restriction of the module $E^{\sigma}$ to the open
subset $\mbox{Spec } {\bf C}[\tau^{\vee}] \hookrightarrow
\mbox{Spec } {\bf C}[\sigma^{\vee}]$.
For consistency, if $\tau$ is a subcone of $\sigma_1$, $\sigma_2$,
then the restrictions of $E^{\sigma_1}$, $E^{\sigma_2}$ must agree.
(This is how modules over overlapping open sets are glued together.)

So far all we have done is define a sheaf (or, rather, a presheaf)
by associating modules of sections to open sets.  To recover
the description of equivariant reflexive sheaves given earlier,
there are two steps.  First, one shows that for a reflexive sheaf,
the module $E^{\sigma}$ associated to any cone $\sigma$ 
is completely determined by the modules $E^{\alpha}$ associated
to toric divisors $\alpha$ in $\sigma$.  (Thus, to specify
a reflexive sheaf, it suffices to know the modules associated
to the one-dimensional edges of the fan.)
Second, one shows that the modules associated to one-dimensional
edges of the fan are all completely determined by filtrations.
Thus, equivariant reflexive sheaves are specified by associating
a filtration to each toric divisor.

This result -- that equivariant reflexive sheaves are specified by
associating a filtration to each toric divisor -- is sufficiently
important to warrant repetition.  The point of interest is that
codimension one behavior is enough to nail down reflexive sheaves;
we need not go to higher codimension.

Now, we shall go over a few details behind these statements.
The modules one sees in studying equivariant sheaves
all have what is essentially an isotypic decomposition
under the action of the algebraic torus.  This means
that we can specify a module by associating a vector space
to each element of the weight lattice of the algebraic torus.
The vector spaces are all subspaces of one fixed vector space -- 
the fiber of the trivial vector bundle over the open torus orbit.

For example, consider ${\bf C}[x,y]$ as a ${\bf C}[x,y]$-module.
Here we associate a vector subspace of ${\bf C}$ to each
element of the weight lattice of $( {\bf C}^{*} )^2$, as follows:
\begin{center}
\begin{tabular}{c|cccccc}
 & & -1 & 0 & 1 & 2 & \\ \hline
 & & & $\vdots$ & & & \\
2 & & 0 & ${\bf C}$ & ${\bf C}$ & ${\bf C}$ & \\
1 & & 0 & ${\bf C}$ & ${\bf C}$ & ${\bf C}$ & \\
0 & $\cdots$ & 0 & ${\bf C}$ & ${\bf C}$ & ${\bf C}$ & $\cdots$ \\
-1 & & 0 & 0 & 0 & 0 & \\
 & & & $\vdots$ & & & \\
\end{tabular}
\end{center}

In general, for any torsion-free ${\bf C}[\sigma^{\vee}]$-module,
multiplying by an element of ${\bf C}[\sigma^{\vee}]$ induces
inclusions.  In the example above, if we let $E(i_1,i_2)$ denote
the vector space associated with monomial $x^{i_1}y^{i_2}$,
then we have inclusions
\begin{eqnarray*}
E(i_1,i_2) & \hookrightarrow & E(i_1+1, i_2) \\
 & \hookrightarrow & E(i_1, i_2+1)
\end{eqnarray*}

For a somwhat less trivial example, consider the ideal
generated by $(x, xy)$ in ${\bf C}[x,xy,xy^2]$.
As a ${\bf C}[x,xy,xy^2]$-module, it has an isotypic decomposition
\begin{center}
\begin{tabular}{c|cccccc}
 & & 0 & 1 & 2 & 3 & \\ \hline
 & & & $\vdots$ & & & \\
5 & & 0 & 0 & 0 & ${\bf C}$ & \\
4 & & 0 & 0 & 0 & ${\bf C}$ & \\
3 & & 0 & 0 & ${\bf C}$ & ${\bf C}$ & \\
2 & $\cdots$ & 0 & 0 & ${\bf C}$ & ${\bf C}$ & $\cdots$ \\
1 & & 0 & ${\bf C}$ & ${\bf C}$ & ${\bf C}$ & \\
0 & & 0 & ${\bf C}$ & ${\bf C}$ & ${\bf C}$ & \\
-1 & & 0 & 0 & 0 & 0 & \\
 & & & $\vdots$ & & & \\
\end{tabular}
\end{center}
With notation as before, it is easy to check one has inclusions
\begin{eqnarray*}
E(i_1, i_2) & \hookrightarrow & E(i_1+1, i_2) \\
  & \hookrightarrow & E(i_1+1, i_2+1 ) \\
  & \hookrightarrow & E(i_1+1, i_2+2) 
\end{eqnarray*}
As a ${\bf C}[x,xy,xy^2]$-module, it has two generators
(located at $x$, $xy$) and one relation.

What is the geometry behind the example above?
$\mbox{Spec } {\bf C}[x,xy,xy^2]$ = ${\bf C}^2/{\bf Z}_2$,
and so it turns out this module defines a reflexive rank 1 sheaf on
${\bf C}^2/{\bf Z}_2$.  In particular, as the module is
not freely generated, this is an example of a reflexive rank 1 sheaf
which is not a line bundle.

So far we have told you about a particularly convenient
(isotypic) decomposition of the modules appearing in equivariant
sheaves.  It turns out that when the module $E^{\sigma}$
associated to cone $\sigma$ is reflexive, it can be specified
in terms of modules associated to one-dimensional fan edges as
\begin{displaymath}
E^{\sigma}(\chi) \: = \: \bigcap_{\alpha \in | \sigma | }
E^{\alpha}(\chi)
\end{displaymath}
We shall not derive this relation here\footnote{We should mention, however,
that it is formally similar to a standard result on reflexive modules
over noetherian integrally closed domains \cite[chapter 7.4]{bourbaki},
which says that if $M$ is a reflexive module over such a domain, 
then 
\begin{displaymath}
M \: = \: \bigcap_{p} M_p
\end{displaymath}
where the intersection is over all prime ideals of height 1.
}, but see instead \cite{meallen}.

So far we have told you that modules defining reflexive
sheaves are completely determined by modules associated to
one-dimensional edges of the fan; we still need to demonstrate
that modules associated to one-dimensional edges of the fan
are completely determined by a filtration.

Why should a module associated to a neighborhood of a toric
divisor be equivalent to a filtration?
A toric neighborhood of any toric divisor is simply
${\bf C} \times ( {\bf C}^{\times} )^n$, which is
associated to a ring, say $A$,
\begin{displaymath}
A \: = \: {\bf C}[x_1, x_2, x_2^{-1}, x_3, x_3^{-1}, \cdots,
x_n, x_n^{-1} ]
\end{displaymath}
Consider the inclusions generated in any associated torsion-free
module:
\begin{eqnarray*}
E(i_1, \cdots, i_n) & \hookrightarrow & E(i_1, i_2+1, \cdots, i_n) \\
   & \hookrightarrow & E(i_1, i_2-1, \cdots, i_n)
\end{eqnarray*}
Clearly, $E(i_1, \cdots, i_n)$ is independent of $i_2, \cdots, i_n$.
The only nontrivial inclusion is simply
\begin{displaymath}
E(i_1, \cdots, i_n) \: \hookrightarrow \: E(i_1+1, i_2, \cdots, i_n)
\end{displaymath}
Thus, this module is equivalent to a filtration.

\begin{figure}
\centerline{\psfig{file=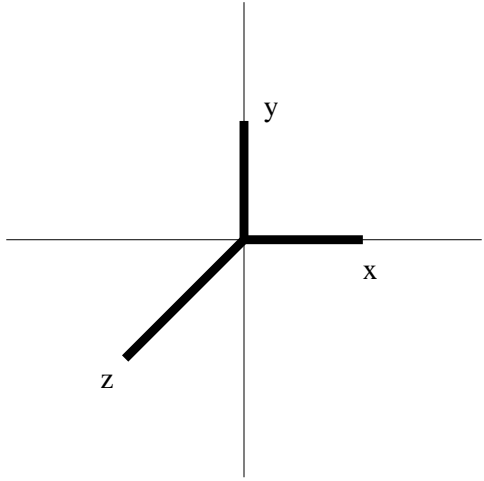,width=2.5in}}
\caption{\label{f1} A fan defining ${\bf P}^2$ as a toric variety}
\end{figure}

Let us consider a simple example to help clarify matters.
Consider the trivial rank 1 line bundle ${\cal O}$ (known more
formally as the structure sheaf)
on ${\bf P}^2$.  Let $x$, $y$, $z$ be homogeneous coordinates
defining ${\bf P}^2$.  A fan defining ${\bf P}^2$ as a toric
variety is shown in figure~\ref{f1}.  The structure sheaf
of a variety is defined by associating to each neighborhood
$\mbox{Spec } A$, a module that is precisely the ring $A$.

First, the toric neighborhood of $D_x = \{ x = 0 \}$ is $\mbox{Spec }
{\bf C}[x,y,y^{-1}]$, and the associated module describing
the structure sheaf is ${\bf C}[x,y,y^{-1}]$:
\begin{center}
\begin{tabular}{c|ccccc}
 & & -1 & 0 & 1 & \\ \hline
 & & & $\vdots$ & & \\
1 & & 0 & ${\bf C}$ & ${\bf C}$ & \\
0 & $\cdots$ & 0 & ${\bf C}$ & ${\bf C}$ & $\cdots$ \\
-1 & & 0 & ${\bf C}$ & ${\bf C}$ & \\
 & & & $\vdots$ & & \\
\end{tabular}
\end{center}

The toric neighborhood of $D_y = \{ y = 0 \}$ is $\mbox{Spec }
{\bf C}[x,x^{-1},y]$, and the associated module describing
the structure sheaf is ${\bf C}[x,x^{-1},y]$:
\begin{center}
\begin{tabular}{c|ccccc}
 & & -1 & 0 & 1 & \\ \hline
 & & & $\vdots$ & & \\
1 & & ${\bf C}$ & ${\bf C}$ & ${\bf C}$ & \\
0 & $\cdots$ & ${\bf C}$ & ${\bf C}$ & ${\bf C}$ & $\cdots$ \\
-1 & & 0 & 0 & 0 & \\
 & & & $\vdots$ & & \\
\end{tabular}
\end{center}

The toric neighborhood of $D_z = \{ z = 0 \}$ is $\mbox{Spec }
{\bf C}[x^{-1} y, x y^{-1}, x^{-1} y^{-1}]$, and the associated
module describing the structure sheaf is 
${\bf C}[x^{-1} y, x y^{-1}, x^{-1} y^{-1}]$:
\begin{center}
\begin{tabular}{c|ccccc}
 & & -1 & 0 & 1 & \\ \hline
 & & & $\vdots$ & & \\
1 & & ${\bf C}$ & 0 & 0 & \\
0 & $\cdots$ & ${\bf C}$ & ${\bf C}$ & 0 & $\cdots$ \\
-1 & & ${\bf C}$ & ${\bf C}$ & ${\bf C}$ & \\
 & & & $\vdots$ & & \\
\end{tabular}
\end{center}

For a reflexive sheaf ${\cal E}$ (of which the structure sheaf 
is a trivial example), if we denote the module associated
to divisor $\alpha$ by $E^{\alpha}$, then the sections of
${\cal E}$ are given by
\begin{displaymath}
H^0({\cal E})_{\chi} \: = \: \bigcap_{\alpha} E^{\alpha}(\chi)
\end{displaymath}
and in particular in the case at hand it is trivial to compute
\begin{displaymath}
H^0( {\bf P}^2, {\cal O} )_{\chi} \: = \:
\left\{ \begin{array}{ll}
        {\bf C} & \chi = 0 \\
        0 & \mbox{otherwise}  \end{array} \right.
\end{displaymath}
so $h^0({\bf P}^2, {\cal O}) = 1$, as is well known.

To summarize, we have argued that for a reflexive sheaf,
the module associated to any cone of the fan is completely
determined by modules associated to one-dimensional fan edges,
and modules associated to one-dimensional fan edges are equivalent
to filtrations.  Thus, to specify an equivariant reflexive sheaf,
we associate a filtration to each toric divisor.

When do two sets of filtrations define the same equivariant
reflexive sheaf?  When they differ by an automorphism of the
topmost vector space.  To construct a moduli space of equivariant
reflexive sheaves, we would have to mod out a space of all filtrations
by automorphisms of the topmost vector space.  In the next section
we shall study such constructions in detail.

\section{Moduli spaces of equivariant sheaves}  \label{modspace}

In this section we will outline how to construct
moduli spaces of equivariant reflexive sheaves via GIT quotients.

First, what is a GIT quotient?
GIT (Geometric Invariant Theory) quotients are closely related to 
symplectic quotients,
and can be loosely thought of as a holomorphic way to compute a 
symplectic quotient \cite[section 8]{git}.
Suppose we have a complex algebraic variety ${\cal T}$
with an action of a reductive algebraic group $G$ and an
ample line bundle ${\cal L}$ on ${\cal T}$.  (The analogous
data for a symplectic quotient would be a symplectic manifold ${\cal T}$
with the action of a compact Lie group $G$ and a specific choice
of symplectic form $\omega$ on ${\cal T}$.)
Points on ${\cal T}$ are classified as stable, semistable, or unstable,
depending upon the action of $G$ and the behavior of ${\cal L}$.
(For example, a point is not stable if the dimension of its
stabilizer in $G$ is greater than zero.)

Denote a GIT quotient by ${\cal T}//G$,
then technically
\begin{displaymath}
{\cal T}//G \: = \: \mbox{Proj } \bigoplus_n H^0 \left(
{\cal T}, {\cal L}^n \right)^G
\end{displaymath}
though for the purposes of this article it will suffice
to say, loosely,
\begin{displaymath}
{\cal T}//G \: = \: \left( {\cal T} - {\cal T}^{us} \right) / G
\end{displaymath}
where ${\cal T}^{us}$ denotes the unstable points of ${\cal T}$.

For example, projective spaces can be realized as very
elementary examples of GIT quotients:
\begin{displaymath}
{\bf P}^n \: = \: {\bf C}^{n+1} // {\bf C}^{\times} \: = \:
\left( {\bf C}^{n+1} - \{ 0 \} \right) / {\bf C}^{\times}
\end{displaymath}

What is the relevant notion of stability for sheaves on a 
K\"ahler variety?  The relevant notion is called ``Mumford-Takemoto
stability.''  (This necessary but not sufficient condition for
a consistent heterotic compactification arises, for example,
as a $D$-term constraint in compactifications to $N=1$ in $3+1$ dimensions.)
For a torsion-free sheaf ${\cal E}$ of rank $r$,
define the slope of ${\cal E}$ to be 
\begin{displaymath}
\mu({\cal E}) \: = \: \frac{c_1({\cal E}) \cup \omega^{n-1}}{r}
\end{displaymath}
where $\omega$ is the K\"ahler form and $n$ is the dimension of the
underlying variety.  With this definition, we say ${\cal E}$
is Mumford-Takemoto (semi)stable if for all proper coherent subsheaves
${\cal F} \subset {\cal E}$ such that $0 < \mbox{rank } {\cal F}
< \mbox{rank } {\cal E}$ and ${\cal E}/{\cal F}$ is torsion-free,
\begin{displaymath}
\mu({\cal F}) \: ( \leq ) < \: \mu({\cal E})
\end{displaymath}

For equivariant reflexive sheaves, the notion of Mumford-Takemoto
stability simplifies.  First, in general for a reflexive
sheaf on any variety it suffices to check only reflexive subsheaves to determine
Mumford-Takemoto stability.  Second, for any equivariant sheaf
it suffices to check only equivariant subsheaves to determine
Mumford-Takemoto stability.  Thus, for equivariant reflexive sheaves,
we need only test equivariant reflexive subsheaves.

Before we can finally construct moduli spaces,
we need a little more information.  Instead of specifying
filtrations, we can specify parabolic\footnote{A parabolic
subgroup of $GL(n,{\bf C})$ is conjugate to a subgroup consisting
of upper-block-triangular matrices.} subgroup of $G = GL(n,{\bf C})$,
as 
\begin{displaymath}
P^{\alpha} \: = \: \{ g \in GL(n,{\bf C}) \, | \,
g E^{\alpha}(i) = E^{\alpha}(i) \, \forall i \}
\end{displaymath}
Specifying a parabolic subgroup $P^{\alpha}$ does not uniquely
identify a filtration $\{ E^{\alpha}(i) \}$ -- it does not
say at which values of $i$ the filtration changes dimension.
That additional information is given by specifying an ample line
bundle on $G/P^{\alpha}$.  Denote this ample line bundle
by ${\cal L}_{\alpha}$.

In terms of parabolics, the constraint for a reflexive sheaf
on a smooth toric variety to be a bundle is that for
all cones $\sigma$,
\begin{equation} \label{constraint}
\bigcap_{\alpha \in | \sigma | } P^{\alpha} \mbox{ contains
a maximal torus of } G
\end{equation}
Any pair of filtrations automatically satisfies this constraint,
so on a smooth toric two-fold, all equivariant reflexive sheaves
are bundles.  (More generally, on a smooth two-dimensional variety,
all reflexive sheaves are bundles -- we have merely noted how this
standard result can be rederived in the equivariant context.)

There is also a description of equivariant principal $G$-bundles
on smooth varieties.  We will not go into much detail,
but the description simply associates a parabolic $P^{\alpha} \subset
G$ and ample line bundle ${\cal L}_{\alpha}$ on $G/P^{\alpha}$
to each toric divisor $\alpha$, satisfying constraint~(\ref{constraint}).

Now we are almost ready to form moduli spaces.
What does the space ${\cal T}$ of equivariant reflexive sheaves
look like before performing a GIT quotient?
Recall to each toric divisor we associated a filtration or
equivalently a parabolic $P^{\alpha}$.  The space of filtrations
of the same form is simply $G/P^{\alpha}$.  Thus, before
quotienting, the space of reflexive sheaves is
\begin{displaymath}
\prod_{\alpha} G/P^{\alpha}
\end{displaymath}
Well, almost.  For nongeneric flags the Chern classes can change,
so truthfully
\begin{displaymath}
{\cal T} \: \subset \: \prod_{\alpha} G/P^{\alpha}
\end{displaymath}
as we want the Chern classes constant on a component of a moduli space.

Finally we can define the relevant GIT quotient.
Recall that two sets of filtrations define the same reflexive sheaf
if they differ by an automorphism of the top vector space,
meaning, if they differ by an element of $G = GL(n,{\bf C})$,
therefore the moduli space we want is simply ${\cal T}//G$,
with ${\cal T}$ constructed as above.
To make sense out of this we must make a specific choice of
ample line bundle on ${\cal T}$.  Let $\pi_{\alpha}: \prod_{\beta} G/P^{\beta}
\rightarrow G/P^{\alpha}$ be the canonical projection,
and let $n_{\alpha} = D_{\alpha} \cup \omega^{n-1}$ be an integer
(for a dense subset of the K\"ahler cone, the $D_{\alpha} \cup
\omega^{n-1}$ will all be proportional to an integer),
then the ample line bundle on ${\cal T}$ is simply
\begin{displaymath}
\otimes_{\alpha} \, \pi_{\alpha}^{*} {\cal L}_{\alpha}^{n_{\alpha}}
\end{displaymath}

In defining this GIT quotient we have implicitly defined some
notion of stability of reflexive sheaves; how does that notion
compare to Mumford-Takemoto stability, the notion of stability
relevant for physics?  It turns out (see \cite{meallen} for details)
that the notion of stability implicit above precisely coincides
with Mumford-Takemoto stability.

In general, GIT quotients of products of flag manifolds are a standard
exercise in the mathematics literature, so in principle a great deal
of information can be extracted from this description.

For example, we can make general remarks concerning singularities
present in such moduli spaces.  Singularities roughly fall
into two classes.  

First, there are singularities present for
nongeneric K\"ahler forms.  As the K\"ahler form is varied, sometimes
semistable sheaves become unstable, or unstable sheaves become
semistable.  When this happens, the topology of the moduli space
changes, and at the transition point there is a singularity.
In extreme cases, such as rank two sheaves on surfaces,
the K\"ahler cone splits into subcones, and one has a topologically
distinct moduli space of sheaves associated to each subcone.
Typically (but not always) these moduli spaces are birational to
one another.  For a review of this phenomenon and 
references in the mathematics literature, see \cite{meallen}.
In general, this sort of behavior of GIT quotients
under change of ample line bundle is ubiquitous; see
\cite{dolgachevhu,thaddeus} for recent expositions.
We shall speak at greater length on this phenomenon in
section~\ref{kcs}.

Secondly, in moduli spaces of principal $G$-bundles (for
$G$ other than $GL(n,{\bf C})$), there are orbifold
singularities, present for generic K\"ahler forms.

\section{More general moduli spaces} \label{genmodspace}

\subsection{More uses of equivariant sheaves}

So far we have only spoken about equivariant sheaves
on toric varieties, though in principle information
can be gained about more general sheaves on 
toric varieties.  

In the mathematics literature, given an action of a
group $G$ on some space, all one needs to know to essentially
reconstruct the space is the fixed points of $G$ and its
action on the normal bundle to the fixed points.
In the present context, given knowledge of equivariant sheaves
and the algebraic torus action on a normal bundle to equivariant
sheaves, one can -- in principle -- reconstruct the rest
of the moduli space.

\subsection{K\"ahler cone substructure}  \label{kcs}

As mentioned previously in section~\ref{modspace},
a necessary but not sufficient condition for a consistent
heterotic compactification is that the sheaf ${\cal E}$ be
Mumford-Takemoto semistable, as defined earlier.
To review, this constraint is satisfied when for all
reasonably well-behaved subsheaves ${\cal F} \subset {\cal E}$,
\begin{displaymath}
\frac{ c_1({\cal F}) \cup \omega^{n-1} }{ \mbox{rank }{\cal F} }
\: \leq \:
\frac{ c_1({\cal E}) \cup \omega^{n-1} }{ \mbox{rank }{\cal E} }
\end{displaymath}
where $\omega$ is the K\"ahler form and $n$ is the dimension
of the variety.  

The relevant point concerning Mumford-Takemoto stability
is that it explicitly depends upon the choice of K\"ahler form.
In particular, as we move around in the K\"ahler cone,
sheaves that are semistable with respect to some K\"ahler
forms may become unstable with respect to others, and vice-versa.

This is an important fact which has so far been completely
overlooked in the physics literature.  At minimum, one can expect
that at certain nongeneric points in the K\"ahler cone, a moduli
space of sheaves will become singular.  More extreme behavior
is also possible.

In particular the case of rank 2 sheaves on complex surfaces
has been thoroughly studied in the mathematics literature
\cite{qin1,qin2,qin3,qin4,friedmanqin,matsukiwentworth,huli,gottsche},
and is closely related to analogous phenomena occuring for
continuous (rather than holomorphic) bundles \cite{friedmanqin2,mooreed}.
For the special case of rank 2 sheaves on complex surfaces,
the K\"ahler cone actually splits into subcones (or ``chambers''),
with a topologically distinct moduli space associated to
each chamber.  (The precise decomposition depends upon the
Chern classes of the sheaves appearing on the moduli space.)
Typically (but not always) the moduli spaces associated to
distinct chambers are birational to one another.

This fact was mentioned previously in section~\ref{modspace},
in the context of equivariant sheaves, where this phenomenon
can be seen explicitly.  However this phenomenon occurs
not only for equivariant sheaves but for general sheaves,
and is sufficiently important to warrant repeating.

\section{(0,2) mirror symmetry} \label{02mirrsec}

There potentially exists a generalization of ordinary mirror
symmetry, known as (0,2) mirror symmetry.

First, recall the definition of ordinary (so-called (2,2) ) mirror
symmetry.  It says that there exist pairs of Calabi-Yaus,
call them $X$, $Y$, both described by the same conformal
field theory -- a string cannot tell which of the two it is
propagating on.

By contrast, (0,2) mirror symmetry exchanges\footnote{In fact,
more complicated examples than this are quite possible,
but for the purposes of this article we shall not go into
such details.} pairs
$(X, {\cal E})$, $(Y, {\cal F})$ where $X$, $Y$ are
Calabi-Yaus and ${\cal E}$, ${\cal F}$ are torsion-free
sheaves on $X$, $Y$, respectively.  (0,2) mirror symmetry
then reduces to ordinary mirror symmetry in the special
case that ${\cal E} = TX$ and ${\cal F} = TY$.

Ordinary mirror symmetry exchanges complex and K\"ahler moduli.
By contrast, (0,2) mirror symmetry is believed to exchange
complex, K\"ahler, and sheaf moduli as a unit:  sheaf moduli may
be mirror not only to other sheaf moduli but perhaps also
to complex or K\"ahler moduli, for example.

In addition, (0,2) mirror symmetry presumably acts on
charged matter as well as neutral moduli. 
Recall ordinary mirror symmetry exchanges Hodge numbers;
for example, for threefolds,
\begin{eqnarray*}
h^{1,1}(X) & = & h^{2,1}(Y) \\
H^{2,1}(X) & = & h^{1,1}(Y)
\end{eqnarray*}
Analogously, at least in simple cases (0,2) mirror symmetry is
believed to exchange global Ext groups
\begin{eqnarray*}
\mbox{Ext}^1_X ( {\cal O}, {\cal E} ) & \cong & \mbox{Ext}^1_Y (
{\cal F}, {\cal O} ) \\
\mbox{Ext}^1_X ( {\cal E}, {\cal O} ) & \cong & \mbox{Ext}^1_Y (
{\cal O}, {\cal F} )
\end{eqnarray*}
In particular, in heterotic compactifications massless modes
are counted by Ext groups \cite{me1,ralph1},
so if ${\cal E}$ and ${\cal F}$ are both rank 3, each embedded
in an $E_8$, then the congruence above simply says that
${\bf 27}$'s and $\overline{{\bf 27}}$'s of $E_6$ are exchanged.
Note that in the special case that ${\cal E} = TX$
we have
\begin{eqnarray*}
\mbox{Ext}^1_X({\cal O}, {\cal E}) & \cong & H^{2,1}(X) \\
\mbox{Ext}^1_Y({\cal E}, {\cal O}) & \cong & H^{1,1}(X) 
\end{eqnarray*} 
and so we recover the analogous expressions for ordinary
mirror symmetry.

Ordinary mirror symmetry is not deeply understood,
but a lot of empirical facts about it are known.
By contrast, very little is known about (0,2) mirror
symmetry \cite{ralph2,ralph3,ralph4}. 
Existing work on the subject has attempted to construct
(0,2) mirrors by orbifolds.  Using such ideas,
one can argue (somewhat weakly) that mirrors to
(restrictions to Calabi-Yaus of) equivariant sheaves
are other equivariant sheaves \cite{meallen}.
An attempt to get insight into how the monomial-divisor
mirror map might be generalized has appeared in \cite{meagain}.

\section{Conclusions}

In this paper we have reviewed a new description of sheaves
on Calabi-Yaus, explained in detail recently in \cite{meallen}
which builds upon work largely done by A. A. Klyachko.
We have also commented on how this work is related to
understanding moduli spaces of more general sheaves,
and on a potential generalization of mirror symmetry.

One of the biggest outstanding problems in string compactifications
is understanding quantum effects in heterotic compactifications;
hopefully our work will be of use in studying this issue.

\section{Acknowledgements}

We would like to thank R. Friedman, T. Gomez, S. Kachru, J. Morgan,
E. Silverstein, K. Uhlenbeck, and E. Witten for useful discussions.
We would also like to thank A. A. Klyachko for supplying us with the
preprints \cite{klyachko3,klyachko4}.


\begin{thebibliography}{199}

\bibitem{meallen} A. Knutson and E. Sharpe, ``Sheaves on Toric
Varieties for Physics,'' {\tt hep-th/9711036}.

\bibitem{klyachko1} A. A. Klyachko, ``Toric Bundles and Problems
of Linear Algebra,'' Funct. Anal. Appl. {\bf 23} (1989) 135.

\bibitem{klyachko2} A. A. Klyachko, ``Equivariant Bundles on 
Toral Varieties,'' Math. USSR Izvestiya {\bf 35} (1990) 337.

\bibitem{klyachko3} A. A. Klyachko, ``Stable Bundles, 
Representation Theory and Hermitian Operators," preprint.

\bibitem{klyachko4} A. A. Klyachko, ``Vector Bundles and Torsion
Free Sheaves on the Projective Plane," preprint.

\bibitem{kaneyama1} T. Kaneyama, ``On Equivariant Vector Bundles
on an Almost Homogeneous Variety," Nagoya Math. J. {\bf 57}
(1975) 65.

\bibitem{kaneyama2} T. Kaneyama, ``Torus-Equivariant Vector
Bundles on Projective Spaces," Nagoya Math. J. {\bf 111} (1988) 25.

\bibitem{fulton} W. Fulton, {\it Introduction to Toric Varieties},
Princeton University Press, 1993.

\bibitem{oda} T. Oda, {\it Convex Bodies and Algebraic Geometry},
Springer-Verlag, 1985.

\bibitem{kkmsd} G. Kempf, F. Knudsen, D. Mumford, and B. Saint-Donat,
{\it Toroidal Embeddings I \/}, Lecture Notes in Mathematics
{\bf 339}, Springer-Verlag, 1973.

\bibitem{danilov} V. I. Danilov, ``The Geometry of Toric Varieties,"
Russian Math Surveys {\bf 33}:2 (1978) 97.


\bibitem{cox} D. A. Cox, ``The Homogeneous Coordinate Ring of
a Toric Variety,'' J. Algebraic Geom. {\bf 4} (1995) 17,
{\tt alg-geom/9210008}.




\bibitem{me1} E. Sharpe, ``Notes on Heterotic Compactifications,"
PUPT-1729, {\tt hep-th/9710031}.

\bibitem{ralph1} R. Blumenhagen, ``(0,2) Target Space Duality,
CICYs and Reflexive Sheaves," {\tt hep-th/9710021}.


\bibitem{bourbaki} N. Bourbaki, {\it Commutative Algebra},
Springer-Verlag, Berlin, 1989.

\bibitem{dolgachevhu} I. Dolgachev and Y. Hu, ``Variation of
Geometric Invariant Theory Quotients,'' {\tt alg-geom/9402008}.

\bibitem{thaddeus} M. Thaddeus, ``Geometric Invariant Theory
and Flips,'' J. Amer. Math. Soc. {\bf 9} (1996) 691.

\bibitem{qin1} Z. Qin, ``Chamber Structures of Algebraic Surfaces
with Kodaira Dimension Zero and Moduli Spaces of Stable Rank Two
Bundles,'' Math. Z. {\bf 207} (1991) 121.

\bibitem{qin2} Z. Qin, ``Birational Properties of Moduli Spaces
of Stable Locally Free Rank-2 Sheaves on Algebraic Surfaces,''
Manuscripta Math. {\bf 72} (1991) 163.

\bibitem{qin3} Z. Qin, ``Equivalence Classes of Polarizations
and Moduli Spaces of Sheaves,'' J. Diff. Geom. {\bf 37} (1993) 397.

\bibitem{qin4} Z. Qin, ``Moduli of Stable Sheaves on Ruled Surfaces
and their Picard Groups,'' J. reine angew. Math. {\bf 433} (1992) 201.

\bibitem{friedmanqin} R. Friedman and Z. Qin, ``On Complex Surfaces
Diffeomorphic to Rational Surfaces,'' Invent. Math. {\bf 120} (1995) 81.

\bibitem{matsukiwentworth} K. Matsuki and R. Wentworth, ``Mumford-Thaddeus
Principle on the Moduli Space of Vector Bundles on an Algebraic Surface,''
Int. J. Math. {\bf 8} (1997) 97.

\bibitem{huli} Y. Hu and W. Li, ``Variation of the Gieseker and Uhlenbeck
Compactifications,'' Int. J. Math. {\bf 6} (1995) 397.

\bibitem{gottsche} L. G\"ottsche, ``Change of Polarization and Hodge
Numbers of Moduli Spaces of Torsion Free Sheaves on Surfaces,''
Math. Z. {\bf 223} (1996) 247.

\bibitem{friedmanqin2} R. Friedman and Z. Qin, ``Flips of Moduli Spaces
and Transition Formulas for Donaldson Polynomial Invariants of
Rational Surfaces,'' Comm. Anal. Geom. {\bf 3} (1995) 11.

\bibitem{mooreed} G. Moore and E. Witten, ``Integration over the
$u$-Plane in Donaldson Theory,'' {\tt hep-th/9709193}.

\bibitem{fmw} R. Friedman, J. Morgan, and E. Witten, ``Vector
Bundles and F theory,'' {\tt hep-th/9701162}.

\bibitem{02} J. Distler and S. Kachru, ``(0,2) Landau-Ginzburg
Theory,'' Nuc. Phys. {\bf B413} (1994) 213, {\tt hep-th/9309110};
J. Distler, ``Notes on (0,2) Superconformal Field Theories,''
{\tt hep-th/9502012}.

\bibitem{ralph2} R. Blumenhagen, R. Schimmrigk, and A. Wisskirchen,
``(0,2) Mirror Symmetry,'' Nucl. Phys. {\bf B486} (1997) 598,
{\tt hep-th/9609167}.

\bibitem{ralph3} R. Blumenhagen and S. Sethi, ``On Orbifolds of
(0,2) Models,'' Nucl. Phys. {\bf B491} (1997) 263, 
{\tt hep-th/9611172}.

\bibitem{ralph4} R. Blumenhagen and M. Flohr, ``Aspects of
(0,2) Orbifolds and Mirror Symmetry,'' Phys. Lett. {\bf B404} (1997) 41,
{\tt hep-th/9702199}.

\bibitem{git} D. Mumford, J. Fogarty, and F. Kirwan, {\it Geometric
Invariant Theory}, third edition, Springer-Verlag, 1994.

\bibitem{meagain} E. Sharpe, ``(0,2) Mirror Symmetry,'' PUPT-1785,
{\tt hep-th/9804066}.


\end{thebibliography}
\end{document}